\documentclass[preprint,aps,showpacs]{revtex4}
\usepackage{graphicx}

\usepackage{amsmath}
\usepackage{amsxtra}
\usepackage{amstext}
\usepackage{amssymb}
\usepackage{latexsym}
\usepackage{dsfont}

\begin{document}

\title{Predictions for fermion masses and mixing from a low energy $SU(3)$ flavor symmetry model with
a Light Sterile Neutrino }

\author{Albino Hern\'andez-Galeana}

\affiliation{ Departamento de F\'{\i}sica,   Escuela Superior de
F\'{\i}sica y Matem\'aticas, I.P.N., \\
U. P. "Adolfo L\'opez Mateos". C. P. 07738, M\'exico, D.F.,
M\'exico. }

\begin{abstract}

I report low energy results on the study of fermion masses and mixing
for quarks and leptons, including neutrinos within a $SU(3)$ flavor symmetry
model, where ordinary heavy fermions, top and bottom quarks and tau lepton become
massive at tree level from {\bf Dirac See-saw} mechanisms
implemented by the introduction a new set of $SU(2)_L$ weak singlet vector-like
fermions $U,D,E,N$, with $N$ a sterile neutrino. Light fermions obtain masses from one loop
radiative corrections mediated by the massive $SU(3)$ gauge
bosons. Recent results shows the existence of a
low energy space parameter where this model is able to accommodate the
known spectrum of quark masses and mixing in a $4\times 4$
non-unitary $V_{CKM}$ as well as the charged lepton masses. Motivated
by the recent LSND and MiniBooNe short-baseline neutrino oscillation
experiments we fit for the 3+1 scenario the neutrino squared mass differences
$m_2^2 - m_1^2\approx 7.6\times 10^{-5}\;\text{eV}^2$,
$m_3^2 - m_2^2\approx 2.43\times 10^{-3}\;\text{eV}^2$ and
$m_4^2 - m_1^2\approx 0.29\;\text{eV}^2$. The model
predicts the D vector like quark mass in the range
$M_D=(\:350\; - \;900\:)\:GeV$ and horizontal
gauge boson masses of few TeV. These low energy predictions are
within LHC possibilities. Furthermore, the above scenario enable us
to suppress simultaneously the tree level $\Delta F=2$ processes for
$K^o-\bar{K^o}$ and $D^o-\bar{D^o}$ meson mixing mediated by these
extra horizontal gauge bosons within current experimental bounds.

\end{abstract}

\keywords{Fermion masses and mixing, Flavor symmetry, Dirac See-saw mechanisms}

\pacs{14.60.Pq, 12.15.Ff, 12.60.-i}
\maketitle

\tableofcontents

%\pagebreak

\section{ Introduction }

The strong hierarchy of quark and charged lepton masses and quark mixing
have suggested to many model building
theorists that light fermion masses could be generated from
radiative corrections\cite{earlyradm}, while those of the top and
bottom quarks as well as that of the tau lepton are generated at
tree level. This may be understood as a consequence of the
breaking of a symmetry among families ( a horizontal symmetry ).
This symmetry may be discrete \cite{modeldiscrete}, or continuous,
\cite{modelcontinuous}. The radiative generation of the light
fermions may be mediated by scalar particles as it is proposed,
for instance, in references \cite{modelrad,medscalars} and this
author in \cite{prd2007}, or also through vectorial bosons as it
happens for instance in "Dynamical Symmetry Breaking" (DSB) and
theories like " Extended Technicolor " \cite{DSB}.

\vspace{2mm}

In this article I address the problem of fermion masses and
quark mixing within an extension of the SM introduced by the
author\cite{albinosu32004} which includes a $SU(3)$ gauged flavor
symmetry commuting with the SM group. In previous
reports\cite{albinosu32009} we showed that this model has the
ingredients to accommodate a realistic spectrum of charged fermion
masses and quark mixing. We introduce a hierarchical mass
generation mechanism in which the light fermions obtain masses
through one loop radiative corrections, mediated by the massive
bosons associated to the $SU(3)$ family symmetry that is
spontaneously broken, while the masses for the top and bottom
quarks as well as for the tau lepton, are generated at tree level
by the implementation of "Dirac See-saw" mechanisms implemented by
the introduction of a new generation of $SU(2)_L$ weak singlets
vector-like fermions. Recently, some authors have pointed out
interesting features regarding the possibility of the existence of
a sequential fourth generation\cite{fourthge}. Theories and models with extra
matter may also provide interesting scenarios for present cosmological problems,
such as candidates for the nature of the Dark Matter
(\cite{normaapproach},\cite{khlopov}). This is the case of an extra
generation of vector-like matter, both from theory and current
experiments\cite{vector-like-SU(2)-weak-singlets}. Due to the fact that the
vector-like quarks do not couple to the $W$ boson, the mixing of $U$ and $D$
vector-like quarks with the SM quarks yield an extended $4\times
4$ non-unitary CKM quark mixing matrix. It has pointed out for
some authors that these type of vector-like fermions are weakly constrained from
Electroweak Precison Data (EWPD) because they do not break
directly the custodial symmetry, then main experimental
constraints on vector-like matter come from the direct production
bounds and their implications on flavor physics. See ref.
\cite{vector-like-SU(2)-weak-singlets} for further details on
constraints for $SU(2)_L$ singlet vector-like fermions.

\vspace{3mm}

Motivated by recent results from the LSND and MiniBooNe short-baseline neutrino oscillation
experiments many authors are paying special attention to the study of
light sterile neutrinos in the eV-scale to explain the tension in the
interpretation of these data\cite{LSND-MiniBooNe}.

Here we report updated low energy results
which accounts for the known quark and charged lepton
masses and the quark mixing in a non-unitary $( V_{CKM} )_{4\times 4}$.
We also include a fit for neutrino masses within a "Dirac See-saw" mechanism
with a light sterile neutrino of $m_4 \approx 0.54 $ eV.

\section{Model with $SU(3)$ flavor symmetry}

\subsection{Fermion content}

We define the gauge group symmetry $G\equiv SU(3) \otimes G_{SM}$
, where $SU(3)$ is a flavor symmetry among families and
$G_{SM}\equiv SU(3)_C \otimes SU(2)_L \otimes U(1)_Y$ is the
"Standard Model" gauge group of elementary particles. The content
of fermions assumes the ordinary quarks and leptons assigned under
G as: $\psi_q^o = ( 3 , 3 , 2 , \frac{1}{3} )_L  \;,\; \psi_l^o =
( 3 , 1 , 2 , -1 )_L \;,\;\psi_u^o = ( 3 , 3, 1 , \frac{4}{3} )_R
\;,\; \psi_d^o = (3, 3 , 1 , -\frac{2}{3} )_R \;,\; \psi_e^o = (3
, 1 , 1,-2)_R $, where the last entry corresponds to the
hypercharge $Y$, and the electric charge is defined by $Q = T_{3L}
+ \frac{1}{2} Y$. The model also includes two types of extra
fermions: Right handed neutrinos $\Psi_\nu^o = ( 3 , 1 , 1 , 0
)_R$, and the $SU(2)_L$ singlet vector-like fermions

 \begin{eqnarray}
U_{L,R}^o= ( 1 , 3 , 1 , \frac{4}{3} )  \qquad , \qquad D_{L,R}^o
= ( 1 , 3 , 1 ,- \frac{2}{3} )  \label{vectorquarks} \\
N_{L,R}^o= ( 1 , 1 , 1 , 0 )
\qquad , \qquad E_{L,R}^o= ( 1 , 1 , 1 , -2 ) \label{vectorleptons}\end{eqnarray}

The above fermion content and its assignment under the group G
make the model anomaly free. After the definition of the gauge
symmetry group and the assignment of the ordinary fermions in the
canonical form under the standard model group and in the
fundamental $3$-representation under the $SU(3)$ family symmetry,
the introduction of the right-handed neutrinos is required to
cancel anomalies\cite{T.Yanagida1979}. The $SU(2)_L$ weak singlets
vector-like fermions have been introduced to give masses at tree
level only to the third family of known fermions through Dirac
See-saw mechanisms. These vector like fermions play a crucial role
to implement a hierarchical spectrum for quarks and charged lepton
masses together with the radiative corrections.

\section{Spontaneous Symmetry breaking}

 The "Spontaneous Symmetry Breaking" (SSB) is proposed to be achieved in the
form:

\begin{equation} G \stackrel{\Lambda_1}{\longrightarrow} SU(2)\otimes G_{SM}
\stackrel{\Lambda_2}{\longrightarrow} G_{SM}
\stackrel{\Lambda_3}{\longrightarrow} SU(3)_C \otimes U(1)_Q  \label{scalesSSB} \end{equation}

\vspace{1mm} \noindent Here $\Lambda_1$, $\Lambda_2$ and $\Lambda_3$ are the scales of
SSB in order the model to have the possibility to be consistent with the known low energy physics.

\subsection{Electroweak symmetry breaking}

To achieve the spontaneous breaking of the electroweak symmetry to
$U(1)_Q$,  we introduce the scalars: $\Phi = ( 3 , 1 , 2 , -1 )$
and $\Phi^{\prime} = ( 3 , 1 , 2 , +1 )$, with the VEV´s: $\langle
\Phi \rangle^T = ( \langle \Phi_1 \rangle , \langle \Phi_2 \rangle
, \langle \Phi_3 \rangle )$ , $\langle \Phi^{\prime} \rangle^T = (
\langle \Phi^{\prime}_1 \rangle , \langle \Phi^{\prime}_2 \rangle
, \langle \Phi^{\prime}_3 \rangle )$, where $T$ means transpose,
and

\begin{equation} \qquad \langle \Phi_i \rangle = \frac{1}{\sqrt[]{2}} \left(
\begin{array}{c} v_i
\\ 0  \end{array} \right) \qquad , \qquad
\langle \Phi^{\prime}_i \rangle = \frac{1}{\sqrt[]{2}} \left(
\begin{array}{c} 0
\\ V_i  \end{array} \right) \:.\end{equation}

\noindent Assuming $(v_1, v_2, v_3) \neq (V_1, V_2, V_3)$ with
$v_1^2+v_2^2+v_3^2=V_1^2+V_2^2+V_3^2 $, the contributions from
$\langle \Phi \rangle$ and $\langle \Phi^{\prime} \rangle$ yield
the $W$ gauge boson mass $\frac{1}{2} g^2 (v_1^2+v_2^2+v_3^2)
W^{+} W^{-} $. Hence, if we define as usual $M_W=\frac{1}{2} g v$,
we may write $ v=\sqrt{2} \sqrt{v_1^2+v_2^2+v_3^2} \thickapprox
246$ GeV.

\vspace{3mm}
Let me emphasize here that solutions for fermion masses and mixing reported
in section \ref{numericalresults} suggest that the dominant contribution to Electroweak Symmetry Breaking
comes from the weak doublets which couple to the third family.

\subsection{$SU(3)$ flavor symmetry breaking}

To implement a hierarchical spectrum for charged fermion masses,
and simultaneously to achieve the SSB of $SU(3)$, we introduce the
scalar fields: $\eta_i,\;i=1,2,3$, transforming under the gauge
group as $(3 , 1 , 1 , 0)$ and taking the "Vacuum Expectation
Values" (VEV's):

\begin{equation} \langle \eta_3 \rangle^T = ( 0 , 0,
{\cal{V}}_3) \quad , \quad \langle \eta_2 \rangle^T = ( 0 ,
{\cal{V}}_2,0) \quad , \quad \langle \eta_1 \rangle^T = (
{\cal{V}}_1,0,0) \:. \end{equation}

\noindent The above scalar fields and VEV's break completely the
$SU(3)$ flavor symmetry. The corresponding $SU(3)$ gauge bosons
are defined in Eq.(\ref{SU3lagrangian}) through their couplings to
fermions. To simplify computations, we impose a $SU(2)$ global
symmetry in the gauge boson masses. So, we assume ${\cal{V}}_1={\cal{V}}_2 \equiv
{\cal{V}}$ in order to cancel mixing between $Z_1$ and $Z_2$
horizontal gauge bosons. Thus, a natural hierarchy among the VEV´s
consistent with the proposed sequence of SSB in Eq.(\ref{scalesSSB}) is
$ {\cal{V}}_3\:>>\:{\cal{V}} \; \gg
\;\sqrt{v_1^2+v_2^2+v_3^2}=\frac{v}{\sqrt{2}}\simeq
\frac{246\:\text{GeV}}{\sqrt{2}} \backsimeq 173.9 \:\text{GeV}
\approx \;m_t $. Hence, neglecting tiny contributions from
electroweak symmetry breaking, we obtain for the gauge bosons
masses

\begin{multline} g_H^2 \left\{ \frac{1}{2} ({\cal{V}})^2
[\:Z_1^2+(Y_1^1)^2+(Y_1^2)^2\:] + \frac{1}{6}\:[\:2
({\cal{V}}_3)^2+({\cal{V}})^2\:]
 \:Z_2^2          \right. \\ \left. + \frac{1}{4}
(\:({\cal{V}}_3)^2+({\cal{V}})^2\:) [\:(Y_2^1)^2+(Y_2^2)^2
+(Y_3^1)^2+(Y_3^2)^2\:] \right\}
\end{multline}

\noindent Them, we may define the horizontal boson masses
\begin{equation}
\begin{array}{rcl}(M_{Z_1})^2=(M_{Y_1^1})^2=(M_{Y_1^2})^2 & = & M_1^2
\equiv g_H^2 {{\cal{V}}}^2  \:,\\
(M_{Y_2^1})^2=(M_{Y_2^2})^2=(M_{Y_3^1})^2=(M_{Y_3^2})^2 & = & M_2^2
\equiv \frac{g_H^2}{2} ({{\cal{V}}_3}^2+{{\cal{V}}}^2 )   \\
(M_{Z_2})^2 & = & 4/3 M_2^2 - 1/3 M_1^2 \end{array} \:, \label{horizontalmasses} \end{equation}

\noindent with the hierarchy $ M_{Z_2} \gtrsim M_2 > M_1 \gg M_W$. It is worth to
 emphasize that this $SU(2)$ global symmetry together with the hierarchy of scales in the SSB
yield a spectrum of $SU(3)$ gauge boson masses without mixing
in quite good approximation. Actually this global $SU(2)$ symmetry plays
the role of a custodial symmetry to suppress properly the tree level
$\Delta F=2$ processes mediated by the $M_1$ lower scale $Z_1,\:Y_1^1,\:Y_1^2$
horizontal gauge bosons.

\section{ Fermion masses}

\subsection{Dirac See-saw mechanisms}

Now we describe briefly the procedure to get the masses for
fermions. The analysis is presented explicitly for the charged
lepton sector, with a completely analogous procedure for the $u$
and $d$ quarks and Dirac neutrinos. With the fields of particles introduced in
the model, we may write the gauge invariant Yukawa couplings, as

\begin{equation}
h\:\bar{\psi}_l^o \:\Phi^\prime \:E_R^o \;+\; h_1 \:\bar{\psi}_e^o \:\eta_1 \:E_L^o \;+\;
h_2 \:\bar{\psi}_e^o \:\eta_2 \:E_L^o \;+\; h_3 \:\bar{\psi}_e^o
\:\eta_3 \:E_L^o \;+\; M \:\bar{E}_L^o \:E_R^o \;+
h.c \label{DiracYC} \end{equation}

\noindent where $M$ is a free mass parameter ( because its mass
term is gauge invariant) and $h$, $h_1$, $h_2$ and $h_3$ are
Yukawa coupling constants. When the involved scalar fields acquire
VEV's we get, in the gauge basis ${\psi^{o}_{L,R}}^T = ( e^{o} ,
\mu^{o} , \tau^{o}, E^o )_{L,R}$, the mass terms $\bar{\psi}^{o}_L
{\cal{M}}^o \psi^{o}_R + h.c $, where

\begin{equation} {\cal M}^o = \begin{pmatrix} 0 & 0 & 0 & h \:v_1\\ 0 & 0 & 0 & h \:v_2\\
0 & 0 & 0 & h \:v_3\\ h_1 {\cal{V}} & h_2 {\cal{V}} & h_3
{\cal{V}}_3 & M \end{pmatrix} \equiv \begin{pmatrix} 0 & 0 & 0 & a_1\\ 0 & 0 & 0 & a_2\\
0 & 0 & 0 & a_3\\ b_1 & b_2 & b_3 & c
\end{pmatrix} \;. \end{equation}

\noindent Notice that ${\cal{M}}^o$ has the same structure of a
See-saw mass matrix, here for Dirac fermion masses.
So, we call ${\cal{M}}^o$ a {\bf "Dirac See-saw"} mass matrix.
${\cal{M}}^o$ is diagonalized by applying a biunitary
transformation $\psi^{o}_{L,R} = V^{o}_{L,R} \;\chi_{L,R}$. The
orthogonal matrices $V^{o}_L$ and $V^{o}_R$ are obtained
explicitly in the Appendix A. From $V_L^o$ and $V_R^o$, and using
the relationships defined in this Appendix, one computes

\begin{eqnarray}
{V^{o}_L}^T {\cal{M}}^{o} \;V^{o}_R =Diag(0,0,-
\sqrt{\lambda_-},\sqrt{\lambda_+})   \label{tleigenvalues}\\
                                  \nonumber   \\
{V^{o}_L}^T {\cal{M}}^{o} {{\cal{M}}^{o}}^T \;V^{o}_L = {V^{o}_R}^T
{{\cal{M}}^{o}}^T {\cal{M}}^{o} \;V^{o}_R =
Diag(0,0,\lambda_-,\lambda_+)  \:.\label{tlLReigenvalues}\end{eqnarray}

\noindent where $\lambda_-$ and $\lambda_+$ are the nonzero
eigenvalues defined in Eqs.(\ref{nonzerotleigenvalues}-\ref{paramtleigenvalues}),
$\sqrt{\lambda_+}$ being the fourth heavy fermion mass, and $\sqrt{\lambda_-}$ of
the order of the top, bottom and tau mass for u, d and e fermions, respectively.
We see from Eqs.(\ref{tleigenvalues},\ref{tlLReigenvalues}) that at tree level the
See-saw mechanism yields two massless eigenvalues associated to the light fermions:

\subsection{One loop contribution to fermion masses}

\begin{figure}[htp] \begin{center}
\includegraphics{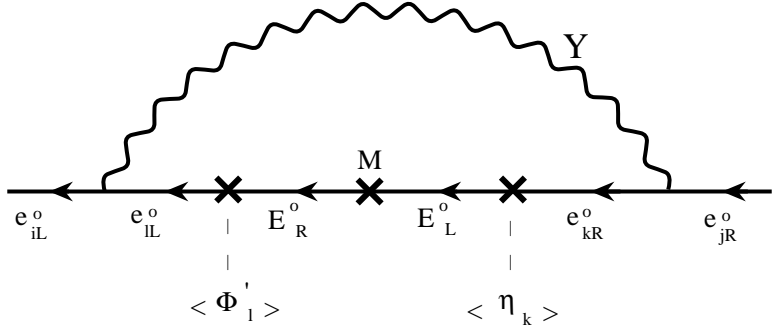} \end{center}
\vspace{-7.5cm}
\caption{ Generic one loop diagram contribution to the mass term
$m_{ij} \:{\bar{e}}_{iL}^o e_{jR}^o$} \end{figure}

Subsequently, the masses for the light fermions arise through one
loop radiative corrections. After the breakdown of the electroweak
symmetry we can construct the generic one loop mass diagram of
Fig. 1. The vertices in this diagram read from the $SU(3)$ flavor
symmetry interaction Lagrangian

\begin{multline} i {\cal{L}}_{int} = \frac{g_{H}}{2} \left\{ (\bar{e^{o}}
\gamma_{\mu} e^{o}- \bar{\mu^{o}} \gamma_{\mu} \mu^{o})Z_1^\mu +
\frac{1}{\sqrt{3}}(\bar{e^{o}} \gamma_{\mu} e^{o}+ \bar{\mu^{o}}
\gamma_{\mu} \mu^{o} - 2 \bar{\tau^{o}}
\gamma_{\mu} \tau^{o})Z_2^\mu    \right.              \\
+ (\bar{e^{o}} \gamma_{\mu} \mu^{o}+ \bar{\mu^{o}} \gamma_{\mu}
e^{o})Y_1^{1 \mu}+(-i\bar{e^{o}} \gamma_{\mu}
\mu^{o}+ i\bar{\mu^{o}} \gamma_{\mu} e^{o})Y_1^{2 \mu}   \\
+(\bar{e^{o}} \gamma_{\mu} \tau^{o}+ \bar{\tau^{o}} \gamma_{\mu}
e^{o})Y_2^{1 \mu}+(-i\bar{e^{o}} \gamma_{\mu}
\tau^{o}+ i\bar{\tau^{o}} \gamma_{\mu} e^{o})Y_2^{2 \mu}  \\
+ \left. (\bar{\mu^{o}} \gamma_{\mu} \tau^{o}+ \bar{\tau^{o}}
\gamma_{\mu} \mu^{o})Y_3^{1 \mu}+(-i\bar{\mu^{o}} \gamma_{\mu}
\tau^{o}+ i\bar{\tau^{o}} \gamma_{\mu} \mu^{o})Y_3^{2 \mu}
\right\} \:,\label{SU3lagrangian}\end{multline}

\noindent where $g_H$ is the $SU(3)$ coupling constant, $Z_1$, $Z_2$
and $Y_i^j\;,i=1,2,3\;,j=1,2$ are the eight gauge bosons. The
crosses in the internal fermion line mean tree level mixing, and
the mass $M$ generated by the Yukawa couplings in Eq.(\ref{DiracYC})
after the scalar fields get VEV's. The one loop diagram of Fig. 1
gives the generic contribution to the mass term $m_{ij}
\:{\bar{e}}_{iL}^o e_{jR}^o$

\begin{equation} c_Y \frac{\alpha_H}{\pi} \sum_{k=3,4} m_k^o
\:(V_L^o)_{ik}(V_R^o)_{jk} f(M_Y, m_k^o) \qquad , \qquad \alpha_H
\equiv \frac{g_H^2}{4 \pi} \end{equation}

\noindent  where $M_Y$ is the gauge boson mass, $c_Y$ is a factor
coupling constant, Eq.(\ref{SU3lagrangian}), $m_3^o=-\sqrt{\lambda_-}$ and
$m_4^o=\sqrt{\lambda_+}$ are the See-saw mass eigenvalues,
Eq.(\ref{tleigenvalues}), and $f(x,y)=\frac{x^2}{x^2-y^2}
\ln{\frac{x^2}{y^2}}$. Using the results of Appendix A, we
compute

\begin{equation} \sum_{k=3,4} m_k^o \:(V_L^o)_{ik}(V_R^o)_{jk} f(M_Y,
m_k^o)= \frac{a_i \:b_j \:M}{\lambda_+ - \lambda_-}\:
F(M_Y,\sqrt{\lambda_-},\sqrt{\lambda_+}) \:,\end{equation}

\noindent $i,j=1,2,3$ and $F(M_Y,\sqrt{\lambda_-},\sqrt{\lambda_+})\equiv
\frac{M_Y^2}{M_Y^2 - \lambda_+} \ln{\frac{M_Y^2}{\lambda_+}} -
\frac{M_Y^2}{M_Y^2 - \lambda_-} \ln{\frac{M_Y^2}{\lambda_-}}$. Adding up all
the one loop $SU(3)$ gauge boson contributions, we get the mass terms
$\bar{\psi^{o}_L} {\cal{M}}_1^o  \:\psi^{o}_R + h.c.$,

\vspace{1mm}

\begin{equation} {\cal{M}}_1^o = \left( \begin{array}{ccrc} R_{11} & R_{12} & R_{13}  & 0\\
R_{21} & R_{22} & R_{23} & 0\\ R_{31} & R_{32} & R_{33} & 0\\
0 & 0 & 0 & 0
\end{array} \right) \:\frac{\alpha_H}{\pi}\;,
\end{equation}

\vspace{1mm}

\begin{equation} R_{11}=-\frac{1}{4} F_1 (m_{11} + 2 m_{22}) - \frac{1}{12}
F_{Z_2} m_{11} + \frac{1}{2} F_2 m_{33}  \:,\nonumber\end{equation}
\vspace{1mm}
\begin{equation} R_{22}=-\frac{1}{4} F_1 (2 m_{11} + m_{22}) -
\frac{1}{12} F_{Z_2} m_{22} + \frac{1}{2} F_2 m_{33} \:,\nonumber\end{equation}
\vspace{1mm}
\begin{equation} R_{12}=(\frac{1}{4} F_1 - \frac{1}{12} F_{Z_2}
)m_{12}\quad ,\quad R_{21}=(\frac{1}{4} F_1 - \frac{1}{12} F_{Z_2}
)m_{21} \;,\end{equation} \vspace{1mm}
\begin{equation} R_{33}=\frac{1}{3} F_{Z_2} m_{33} -
\frac{1}{2} F_2 (m_{11} + m_{22}) \quad ,\quad  R_{13}= -
\frac{1}{6} F_{Z_2} m_{13} \:,\nonumber            \end{equation} \vspace{1mm}
\begin{equation} R_{31}= \frac{1}{6} F_{Z_2} m_{31} \quad ,\quad R_{23}= -
\frac{1}{6} F_{Z_2} m_{23}  \quad ,\quad R_{32}= \frac{1}{6} F_{Z_2}
m_{32} \;. \nonumber \end{equation}

\vspace{3mm} \noindent Here, $F_{1,2} \equiv
F(M_{1,2},\sqrt{\lambda_-},\sqrt{\lambda_+})$ and $F_{Z2} \equiv
F(M_{Z_2},\sqrt{\lambda_-},\sqrt{\lambda_+})$, with $M_1 \:,\:M_2$
and $M_{Z_2}$ the horizontal boson masses, Eq.(\ref{horizontalmasses}),

\begin{equation} m_{ij}=\frac{a_i \:b_j \:M}{\lambda_+ - \lambda_-} = \frac{a_i
\:b_j}{a \:b} \:\sqrt{\lambda_-}\:c_{\alpha} \:c_{\beta} \:,\end{equation}

\noindent and $c_{\alpha} \equiv \cos\alpha \:,\;c_{\beta} \equiv \cos\beta \:,\;
s_{\alpha} \equiv \sin\alpha \:,\;s_{\beta} \equiv \sin\beta$, as defined in the
Appendix, Eq.(\ref{Seesawmixing}). Therefore, up to one loop
corrections we obtain the fermion masses

\begin{equation} \bar{\psi}^{o}_L {\cal{M}}^{o} \:\psi^{o}_R + \bar{\psi^{o}_L}
{\cal{M}}_1^o \:\psi^{o}_R = \bar{\chi_L} \:{\cal{M}}
\:\chi_R \:,\end{equation}

\vspace{1mm} \noindent with ${\cal{M}} \equiv  \left[ Diag(0,0,-
\sqrt{\lambda_-},\sqrt{\lambda_+})+ {V_L^o}^T {\cal{M}}_1^o
\:V_R^o \right]$.

\pagebreak

Using $V_L^o$, $V_R^o$ in Eqs.(\ref{VoLVI}-\ref{VoRVI}) we get the mass matrix in
Version I:

\begin{equation} {\cal{M}}= \left( \begin{array}{rrcc} m_{11}&m_{12}&c_\beta \:m_{13}&s_\beta \:m_{13} \\
                                                             \\
m_{21}& m_{22} & c_\beta \:m_{23} & s_\beta \:m_{23}\\
                                                             \\
c_\alpha \:m_{31}& c_\alpha \:m_{32} & (-\sqrt{\lambda_-}+c_\alpha c_\beta
\:m_{33}) & c_\alpha s_\beta \:m_{33} \\
                                           \\
s_\alpha \:m_{31}& s_\alpha \:m_{32} & s_\alpha c_\beta \:m_{33} &
(\sqrt{\lambda_+}+s_\alpha s_\beta \:m_{33})
\end{array} \right) \;,\label{massVI}\end{equation}

\vspace{4mm} \noindent where the mass entries $m_{ij}\: ;i,j=1,2,3$ are written
as:

\begin{equation} \begin{array}{ll}
m_{11}=\frac{\eta_+}{a^\prime b^\prime}  c_1 H\:,&
m_{12}= - \frac{\eta_-}{a^\prime b^\prime } \frac{b_3}{b} c_1 H \:,\\
                                                       \\
m_{21}= \frac{\eta_-}{a^\prime b^\prime } \frac{a_3}{a} c_1 H  \:,&
m_{22}=c_2 \left[ \frac{\eta_+}{a^\prime b^\prime} H + \frac{a^\prime b^\prime}{a_3 b_3 } (J+ \frac{\Delta}{2})     \right] \:, \\
                                                       \\
m_{31}= \frac{\eta_-}{a^\prime b^\prime } \frac{a^\prime}{a} c_1 H   ,&
m_{32}=c_2 \left[ \frac{a^\prime}{a_3} (\frac{\eta_+}{a^\prime b^\prime} H + \frac{1}{2} \frac{a^\prime b^\prime}{a_3 b_3 } \Delta) - \frac{b^\prime}{b_3} J \right] \:, \\
                                                   \\
m_{13}= - \frac{\eta_-}{a^\prime b^\prime } \frac{b^\prime}{b} c_1 H ,& m_{23}=\left[ \frac{b^\prime}{b_3} (\frac{\eta_+}{a^\prime b^\prime} H + \frac{1}{2} \frac{a^\prime b^\prime}{a_3 b_3 } \Delta) - \frac{a^\prime}{a_3} J \right]     \:, \end{array}   \end{equation}

\vspace{4mm}
\begin{equation} m_{33}=c_2 \left( \frac{\eta_+}{a_3 b_3} H + J + \frac{1}{6} \frac{ {a^\prime}^2 {b^\prime}^2}{a_3^2 b_3^2 } \Delta - \frac{1}{3} \left[\frac{ {a^\prime}^2 {b^\prime}^2}{a_3^2 b_3^2 } F_1 + (1+ \frac{{a^\prime}^2}{a_3^2}+ \frac{{b^\prime}^2}{b_3^2} )F_{Z_2}     \right]   \right) \:,\nonumber \end{equation}

\vspace{5mm}

\noindent For $V_L^o$, $V_R^o$ of Eqs.(\ref{VoLVII}-\ref{VoRVII}) we get the Version II:

\begin{equation} {\cal M}= \left( \begin{array}{rrcc} M_{11}&M_{12}&c_\beta \:M_{13}&s_\beta \:M_{13} \\
                                                             \\
M_{21}& M_{22} & c_\beta \:M_{23} & s_\beta \:M_{23}\\
                                                             \\
c_\alpha \:M_{31}& c_\alpha \:M_{32} & (-\sqrt{\lambda_-}+c_\alpha c_\beta
\:M_{33}) & c_\alpha s_\beta \:M_{33} \\
                                           \\
s_\alpha \:M_{31}& s_\alpha \:M_{32} & s_\alpha c_\beta \:M_{33} &
(\sqrt{\lambda_+}+s_\alpha s_\beta \:M_{33})
\end{array} \right) \;,\label{massVII} \end{equation}

\vspace{4mm} \noindent where the mass terms $M_{ij}\: ;i,j=1,2,3$ may be obtained from those of
 $m_{ij}$ as follows

\begin{equation} \begin{array}{ccc}
M_{11}=m_{22}  ,& M_{12}=- m_{21} , & M_{13}=m_{23} \\
                                                   \\
M_{21}=-m_{12} ,& M_{22}=m_{11} , & M_{23}=-m_{13} \\
                                                   \\
M_{31}=m_{32} ,& M_{32}=-m_{31} , & M_{33}=m_{33} \end{array}
\end{equation}

\vspace{3mm}

\begin{equation} \eta_-=a_1 \:b_2 - a_2 \:b_1 \quad , \quad \eta_+=a_1 \:b_1 + a_2
\:b_2 \quad , \quad \eta_-^2 + \eta_+^2= {a^\prime}^2 {b^\prime}^2 \end{equation}

\begin{equation} a^\prime=\sqrt{a_1^2+a_2^2}\;\;, \;\;
b^\prime=\sqrt{b_1^2+b_2^2} \;\;, \;\;a=\sqrt{{a^\prime}^2+a_3^2} \;\; ,
\;\; b=\sqrt{{b^\prime}^2+b_3^2} \;, \end{equation}

\begin{equation} c_1=\frac{1}{2} c_\alpha c_\beta \frac{a_3 \:b_3}{a\:b}
\frac{\alpha_H}{\pi} \quad , \quad c_2= \frac{a_3 \:b_3}{a\:b}
\:c_1   \:, \end{equation}

\begin{equation} H=F_2 + \frac{\eta_+}{a_3 b_3} \:F_1 \quad ,
\quad J=F_{Z_2} + \frac{\eta_+}{a_3 b_3}\:F_2 \quad , \quad \Delta=F_{Z_2} - F_1\:.
\end{equation}

\vspace{4mm}

\noindent The diagonalization of ${\cal{M}}$,
Eq.(\ref{massVI}) or Eq.(\ref{massVII}), gives the physical masses for  u,
d, e and $\nu$ fermions. Using a new biunitary transformation
$\chi_{L,R}=V_{L,R}^{(1)} \;\Psi_{L,R}$;
\;$\bar{\chi}_L \;{\cal{M}} \;\chi_R= \bar{\Psi}_L \:{V_L^{(1)}}^T
{\cal{M}} \; V_R^{(1)} \:\Psi_R $, with ${\Psi_{L,R}}^T = ( f_1 ,
f_2 , f_3 , F )_{L,R}$ the mass eigenfields, that is

\begin{equation}
{V^{(1)}_L}^T {\cal{M}} \:{\cal M}^T \;V^{(1)}_L =
{V^{(1)}_R}^T {\cal M}^T \:{\cal{M}} \;V^{(1)}_R =
Diag(m_1^2,m_2^2,m_3^2,M_F^2) \:,\end{equation}

\noindent $m_1^2=m_e^2$, $m_2^2=m_\mu^2$, $m_3^2=m_\tau^2$ and
$M_F^2=M_E^2$ for charged leptons. Therefore, the transformation from
massless to mass fermions eigenfields in this scenario reads

\begin{equation} \psi_L^o = V_L^{o} \:V^{(1)}_L \:\Psi_L \qquad \mbox{and}
\qquad \psi_R^o = V_R^{o} \:V^{(1)}_R \:\Psi_R \end{equation}

\subsection{Quark Mixing and non-unitary $( V_{CKM} )_{4\times 4}$ }

Recall that vector like quarks, Eq.(\ref{vectorquarks}), are $SU(2)_L$
weak singlets, and then they do not couple to $W$ boson in the
interaction basis. So, the interaction of quarks ${f_{uL}^o}^T=(u^o,c^o,t^o)_L$ and
${f_{dL}^o}^T=(d^o,s^o,b^o)_L$ to the $W$ charged gauge boson
is

\begin{equation} \bar{f^o}_{u L} \gamma_\mu f_{d L}^o
{W^+}^\mu = \bar{\Psi}_{u L}\;{V_{u L}^{(1)}}^T\;[(V_{u
L}^o)_{3\times 4}]^T \;(V_{d L}^o)_{3\times 4} \;V_{d
L}^{(1)}\;\gamma_\mu \Psi_{d L} \;{W^+}^\mu \:,\end{equation}

\noindent hence, the non-unitary $V_{CKM}$ of dimension $4\times
4$ is identified as

\begin{equation} (V_{CKM})_{4\times 4}\equiv {V_{u L}^{(1)}}^T\;[(V_{u
L}^o)_{3\times 4}]^T \;(V_{d L}^o)_{3\times 4} \;V_{d L}^{(1)}
\:.\end{equation}

\noindent For u-quarks in version I and d-quarks in version II,

\begin{equation} V^o \equiv [(V_{u L}^o)_{3\times 4}]^T \;(V_{d L}^o)_{3\times 4} = \begin{pmatrix}
\frac{s_o}{\sqrt{1+r_d^2}} & - c_o & \frac{c_\alpha^d s_o r_d}{\sqrt{1+r_d^2}} & \frac{s_\alpha^d s_o r_d}{\sqrt{1+r_d^2}}  \\
                      \\
\Omega_{11} & \frac{s_o}{\sqrt{1+r_u^2}} & c_\alpha^d \: \Omega_{13} &  s_\alpha^d \: \Omega_{13} \\
                                                                                       \\
c_\alpha^u \: \Omega_{31} & \frac{c_\alpha^u s_o r_u}{\sqrt{1+r_u^2}} & c_\alpha^u \: c_\alpha^d \: \Omega_{33}  & c_\alpha^u \: s_\alpha^d \: \Omega_{33} \\
                                                      \\
s_\alpha^u \: \Omega_{31}   &    \frac{s_\alpha^u s_o r_u}{\sqrt{1+r_u^2}}& s_\alpha^u \: c_\alpha^d \:
\Omega_{33}   & s_\alpha^u \: s_\alpha^d \: \Omega_{33}
\end{pmatrix}  \:, \label{Vo} \end{equation}

\vspace{3mm}

\begin{equation} \Omega_{11}=\frac{r_u r_d + c_o}{\sqrt{(1+r_u^2)(1+r_d^2)}}
\quad , \quad \Omega_{13}=\frac{r_d c_o -
r_u}{\sqrt{(1+r_u^2)(1+r_d^2)}} \end{equation}

\begin{equation} \Omega_{31}=\frac{r_u c_o - r_d}{\sqrt{(1+r_u^2)(1+r_d^2)}}
\quad , \quad  \Omega_{33}=\frac{r_u r_d c_o +
1}{\sqrt{(1+r_u^2)(1+r_d^2)}} \end{equation}

\begin{equation} s_o=\frac{v_2}{v^\prime}\:\frac{V_1}{V^\prime} - \frac{v_1}{v^\prime}\:\frac{V_2}{V^\prime}
\quad , \quad
c_o=\frac{v_1}{v^\prime}\:\frac{V_1}{V^\prime} +
\frac{v_2}{v^\prime}\:\frac{V_2}{V^\prime} \label{somixing} \end{equation}

\begin{equation} c_o^2+s_o^2=1 \quad , \quad r_u=(\frac{a^\prime}{a_3})_u \quad
, \quad r_d=(\frac{a^\prime}{a_3})_d        \end{equation}

\vspace{2mm} \noindent $V_i , \;v_i\;,i=1,2$ are related to (e,d)
and (u,$\nu$) fermions respectively.

\section{Numerical results}\label{numericalresults}

Using the strong hierarchy for quarks and charged leptons masses
and the results in\cite{prd2007}, we report here the magnitudes of
quark masses and mixing coming from the analysis of a low energy
parameter space in this model. For this numerical analysis
we used the input global parameters $\frac{\alpha_H}{\pi}=0.2$,
$M_1=4$ TeV and $M_2=1700$ TeV.

\vspace{2mm}

\subsection{Sector d:}

Parameter space: $(\sqrt{\lambda_-})_d= 4.98$ GeV,
$(\sqrt{\lambda_+})_d =500$ GeV, $r_d=0.052$, $(\eta_+/{a_3 b_3})_d=-0.49$,
$(\eta_-/\eta_+)_d=1.3$, $s_\alpha^d=0.01$, and $s_\beta^d=0.7056$, lead to
the down quark masses: $m_d=5.4663$ MeV, $m_s=107.699$ MeV, $m_b=4.216$
GeV, {\bf $M_D=500.008$ GeV}, and the mixing matrix

\small
\begin{equation} V_{d L}^{(1)}= \left(
\begin{array}{rrrr}
0.61120 &-0.79139  &-0.01093  &9.2\times 10^{-5} \\
0.79127  &0.61129  &-0.01429 &  1.2 \times 10^{-4}\\
0.01799& 8.04\times 10^{-5} & 0.99983 & 0.00152\\
-1.78\times 10^{-4}& - 7.96 \times 10^{-7}& -0.00152 & 0.99999
\end{array} \right) \:.\label{VdL} \end{equation}
\normalsize

\subsection{Sector u:}

Parameter space: $(\sqrt{\lambda_-})_u= 358.2$ GeV,
$(\sqrt{\lambda_+})_u =1241.44$ TeV, $r_u=.04$, $(\eta_+/{a_3 b_3})_u=-3.20432$,
$( \eta_-/\eta_+)_u=0$, $s_\alpha^u=.01$ and $s_\beta^u=0.02884$ yield the
up quark masses $m_u=2.4$ MeV, $m_c=1.2$ GeV, $m_t=172$ GeV, $M_U=1241.44$ TeV, and
the mixing

\begin{equation} V_{u L}^{(1)}=
\begin{pmatrix}
1 & 0 &0 & 0\\
0 & 0.99900  & 0.04458 &  -1.80 \times 10^{-7} \\
0 &-0.04458 &0.99900 & 4.34 \times 10^{-6} \\
0 & 3.7366. \times 10^{-7} & -4.33 \times 10^{-6} & 1
\end{pmatrix}  \:.\end{equation}

The See-saw $V^o$ contribution, Eq.(\ref{Vo}) with $s_o=-0.417698$, Eq.(\ref{somixing})  reads

\begin{equation} V^o=
\begin{pmatrix}
0.41713  &0.90858  &- 0.02168  &-2.16 \times 10^{-4} \\
- 0.90871  &0.41736  &0.00723 &  7.23 \times 10^{-5} \\
0.01562 & 0.01669 &0.99963 & 0.01\\
1.562 \times 10^{-4}& 1.66 \times 10^{-4}& 0.0100 & 0.0001
\end{pmatrix} \end{equation}

\subsection{$(V_{CKM})_{4\times 4}$}

The above up and down quark mixing matrices $V_{u L}^{(1)}$,
$V_{d L}^{(1)}$ and $V^o$ yield the quark mixing matrix

\vspace{4mm}

\begin{equation} (V_{CKM})_{4 \times 4} =
\begin{pmatrix}
0.97428  &0.22530  &0.00413  & -3.97 \times 10^{-4} \\
-0.22527  &0.97341  &0.04133 &  -3.96 \times 10^{-4} \\
0.00528&-0.04120 & 0.99902 & -0.01151\\
4.77 \times 10^{-5}& - 2.25 \times 10^{-5}& -0.0100 & 1.15 \times 10^{-4}
\end{pmatrix} \label{Vckm4x4} \end{equation}

\vspace{4mm}

\noindent Notice that the $(V_{CKM})_{3 \times 3}$ sub-matrix is nearly a
unitary mixing matrix, which is consistent with the allowed measured values for quark mixing
reported in the PDG \cite{PDG2010}.

\subsection{Charged Leptons:}

For this sector, the parameter space: $(\sqrt{\lambda_-})_e=
9.14301$ GeV, $(\sqrt{\lambda_+})_e =23816.4$ TeV, $r_e=0.05$,
$(\eta_+/{a_3 b_3})_e=-1.99484$, $(\eta_-/\eta_+)_e=0$, $s_\alpha^e=0.001$ and
$s_\beta^e=0.00038$, reproduce the known charged lepton masses:
$m_e=0.511$ MeV , $m_\mu=105.658$ MeV, $m_\tau=1776.82$ MeV and $M_E \thickapprox 23816.4$ TeV

\subsection{Neutrinos 3+1:}

For this sector, the parameter space: $(\sqrt{\lambda_-})_\nu=
0.048$ eV, $(\sqrt{\lambda_+})_e =0.54$ eV, $r_\nu=0.04$,
$(\eta_+/{a_3 b_3})_e=0.01$, $(\eta_-/\eta_+)_e=4.7$, $s_\alpha^\nu=0.2$ and
$s_\beta^\nu=0.3992$, fit the neutrinos masses

\begin{equation} (m_1,m_2,m_3,m_4)=(\:0.0102\:,\:0.0134\:,\:0.0511\:,\:0.5398\:)\;eV \;,\end{equation}

\noindent the squared mass differences

\begin{eqnarray} m_2^2 - m_1^2\approx 7.6\times 10^{-5}\;\text{eV}^2  ,&
m_3^2 - m_2^2\approx 2.43\times 10^{-3}\;\text{eV}^2   \nonumber \\
                                               \nonumber   \\
m_4^2 - m_1^2\approx 0.29\;\text{eV}^2 &
\;,\end{eqnarray}

\noindent and for charged leptons and neutrinos in version I,
the first row of lepton mixing angles

\begin{eqnarray} (U_{PMNS})_{11}=0.8145  \;,&  (U_{PMNS})_{12}=0.5773 \nonumber \\
                                                      \nonumber \\
 (U_{PMNS})_{13}=0.0422 \;,& (U_{PMNS})_{14}=1.27\times 10^{-4}
\end{eqnarray}

\subsection{FCNC's in $K^o-\bar{K^o}$ meson mixing}

The $SU(3)$ horizontal gauge bosons contribute to new FCNC's, in
particular they mediate $\Delta F=2$ processes at tree level. Here
we compute their leading contribution to $K^o-\bar{K^o}$ meson
mixing. In the previous scenario the up quark sector does not
contribute to $(V_{CKM})_{12}$, and hence the effective
hamiltonian from the tree level diagrams, Fig.2, mediated by the
$SU(2)$ horizontal gauge bosons of mass $M_1$ to the ${\cal
O}_{LL}(\Delta S=2)=(\bar{d}_L \gamma_\mu s_L)(\bar{d}_L
\gamma^\mu s_L)$ operator, is given by

\begin{equation} {\cal H }_{eff} = C_{\bar{d} s}\:{\cal O}_{LL} \quad , \quad
C_{\bar{d} s} \approx \frac{g_H^2}{4}  \frac{1}{M_1^2}
\frac{r_d^4}{(1+r_d^2)^2} (s_{12}^d)^2 \:,\end{equation}

\noindent and then contribute to the $K^o-\bar{K^o}$ mass
difference as

\begin{equation} \Delta M_K \approx \frac{2 \pi^2}{3}   \frac{\alpha_H}{\pi}
\frac{r_d^4}{(1+r_d^2)^2} (s_{12}^d)^2 \frac{F_K^2}{M_1^2}
B_K(\mu) M_K \:. \label{DeltaMK} \end{equation}

\noindent It is worth to point out the double mixing suppression in $\Delta M_K $, Eq.(\ref{DeltaMK});
one from the see-saw mechanism due to the $r_d= (\frac{a^\prime}{a_3})_d$ parameter,
and the one from d-quark mixing $s_{12}^d$. Using the input values:
$r_d=0.052$, $\frac{\alpha_H}{\pi} =0.2 $, $s_{12}^d=0.79139$, $F_K=160$ MeV, $M_K=497.614$ MeV and $B_K=0.8$, one gets

\begin{equation} \Delta m_K \thickapprox 2.77 \times {10}^{-12}\:\text{MeV} \; ,
\end{equation}

\noindent which is lower than the current experimental
bound\cite{PDG2010}, $(\Delta m_K)_{Exp} = M_{K_L} - M_{K_S}
\thickapprox 3.48 \times {10}^{-12}\:\text{MeV}$. The quark
mixing alignment in Eqs.(\ref{VdL} - \ref{Vckm4x4}) avoids tree
level contributions to $D^0-\bar{D^o}$ mixing mediated by the
$SU(2)$ horizontal gauge bosons.

\vspace{5mm}

\begin{figure} \begin{center} \includegraphics[width=0.9\columnwidth]{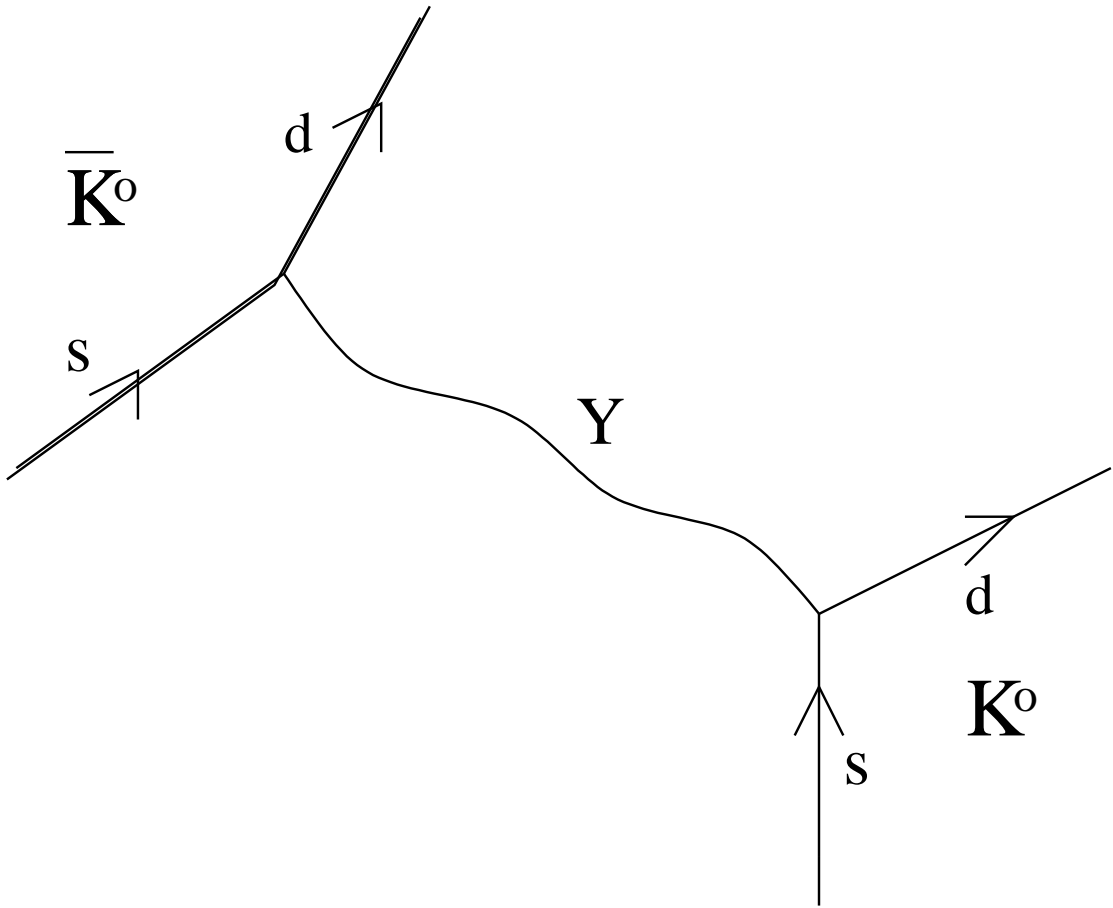}
\caption{Tree level contribution to  $K^o-\bar{K^o}$ from the light $SU(2)$ horizontal gauge bosons.}
\end{center} \end{figure}

\vspace{5mm}

\section{Conclusions}

We have reported a low energy parameter space within a $SU(3)$
flavor symmetry model extension, which combines tree level "Dirac
See-saw" mechanisms and radiative corrections to implement a
successful hierarchical spectrum for charged fermion masses and quark mixing.
In section 5 we have reported the predicted values for quark and charged
lepton masses and quark mixing matrix $(V_{CKM})_{4\times 4}$ within allowed
experimental values reported in PDG 2010, coming from an input
space parameter region with the lower horizontal scale $M_1=4$ TeV
and a D vector-like quark mass of the order of $500$ GeV.
Furthermore, motivated by the recent LSND and MiniBooNe short-baseline neutrino oscillation
experiments we are able to fit in the 3+1 scenario the neutrino squared mass differences
$m_2^2 - m_1^2\approx 7.6\times 10^{-5}\;\text{eV}^2$,
$m_3^2 - m_2^2\approx 2.43\times 10^{-3}\;\text{eV}^2$ and
$m_4^2 - m_1^2\approx 0.29\;\text{eV}^2$. Hence some of the new particles
introduced in this model are within reach at
the current LHC experiments, while simultaneously being consistent with
present bounds on FCNC in $K^o-\bar{K^o}$ and $D^o-\bar{D^o}$ meson mixing.

\vspace{5mm}

\section{Acknowledgments}

It is my pleasure to thank the organizers for invitation, useful discussions and for
the stimulating Workshop at Bled, Slovenia. The author acknowledgment
partial support from the "Instituto Polit\'ecnico Nacional",
(Grants from EDI and COFAA) and "Sistema Nacional de
Investigadores" (SNI) in Mexico.

\section{Appendix A: \\Diagonalization of the generic Dirac See-saw mass matrix}

\begin{equation} {\cal M}^o=
\begin{pmatrix} 0 & 0 & 0 & a_1\\ 0 & 0 & 0 & a_2\\ 0 & 0 & 0 &
a_3\\ b_1 & b_2 & b_3 & c \end{pmatrix} \end{equation}

\vspace{1mm} \noindent Using a biunitary transformation
$\psi^{o}_L = V_L^o \:\chi_L$ and  $\psi^{o}_R = V_R^o
\:\chi_R $ to diagonalize ${\cal{M}}^o$, the orthogonal matrices
$V^{o}_L$ and $V^{o}_R$ may be written explicitly as the following
two versions

\vspace{4mm}
Version I:
\vspace{3mm}

\begin{equation} V^{o}_L = \left( \begin{array}{ccrr}\frac{a_2}{a^\prime}& \frac{a_1 a_3}{a^\prime
a} & \frac{a_1}{a} \cos\alpha &
\frac{a_1}{a} \sin\alpha\\
                        \\
- \frac{a_1}{a^\prime}  & \frac{a_2 a_3}{a^\prime a}  &
\frac{a_2}{a} \cos\alpha &
\frac{a_2}{a} \sin\alpha\\
                        \\
0 & - \frac{a^\prime}{a}   & \frac{a_3}{a} \cos{\alpha}
& \frac{a_3}{a} \sin{\alpha}\\
                            \\
0 & 0 & -\sin{\alpha} & \cos{\alpha}
\end{array} \right) \label{VoLVI} \end{equation}

\vspace{3mm}

\begin{equation} V^{o}_R = \left( \begin{array}{ccrr}
\frac{b_2}{b^\prime} & \frac{b_1 b_3}{b^\prime b} & \frac{b_1}{b} \cos{\beta} &
\frac{b_1}{b} \sin{\beta}\\
                         \\
- \frac{b_1}{b^\prime} & \frac{b_2 b_3}{b^\prime b} &
\frac{b_2}{b} \cos{\beta} & \frac{b_2}{b} \sin{\beta}\\
                                                     \\
0& - \frac{b^\prime}{b} & \frac{b_3}{b} \cos{\beta} &
\frac{b_3}{b} \sin{\beta}\\
                         \\
0 & 0 & -\sin{\beta} & \cos{\beta}
\end{array} \right) \label{VoRVI} \end{equation}

\vspace{5mm}

Version II:
\vspace{3mm}

\begin{equation} V^{o}_L = \left( \begin{array}{ccrr} \frac{a_1 a_3}{a^\prime
a}  & - \frac{a_2}{a^\prime}    & \frac{a_1}{a} \cos\alpha &
\frac{a_1}{a} \sin\alpha\\
                        \\
\frac{a_2 a_3}{a^\prime a}  & \frac{a_1}{a^\prime}    &
\frac{a_2}{a} \cos\alpha &
\frac{a_2}{a} \sin\alpha\\
                        \\
- \frac{a^\prime}{a} &   0  & \frac{a_3}{a} \cos{\alpha}
& \frac{a_3}{a} \sin{\alpha}\\
                            \\
0 & 0 & -\sin{\alpha} & \cos{\alpha}
\end{array} \right) \;, \label{VoLVII} \end{equation}

\vspace{3mm}

\begin{equation} V^{o}_R = \left( \begin{array}{ccrr}
\frac{b_1 b_3}{b^\prime b} & - \frac{b_2}{b^\prime} & \frac{b_1}{b} \cos{\beta} &
\frac{b_1}{b} \sin{\beta}\\
                         \\
\frac{b_2 b_3}{b^\prime b} & \frac{b_1}{b^\prime} &
\frac{b_2}{b} \cos{\beta} & \frac{b_2}{b} \sin{\beta}\\
                                                     \\
 - \frac{b^\prime}{b}& 0 & \frac{b_3}{b} \cos{\beta} &
\frac{b_3}{b} \sin{\beta}\\
                         \\
0 & 0 & -\sin{\beta} & \cos{\beta}
\end{array} \right) \;,\label{VoRVII} \end{equation}

\vspace{4mm}

\begin{equation} \lambda_{\pm } = \frac{1}{2} \left( B \pm \sqrt{B^2 -4D} \right) \label{nonzerotleigenvalues}
\end{equation}

\vspace{3mm} \noindent are the nonzero eigenvalues of
${\cal{M}}^{o} {{\cal{M}}^{o}}^T$ (${{\cal{M}}^{o}}^T
{\cal{M}}^{o}$), with

\begin{eqnarray} B = a^2 + b^2 + c^2 =
\lambda_{-}+\lambda_{+}\quad &, \quad D= a^2
b^2=\lambda_{-}\lambda_{+} \;,\label{paramtleigenvalues} \end{eqnarray}

\vspace{1mm}

 \begin{eqnarray} \cos{\alpha} =\sqrt{\frac{\lambda_+ -
a^2}{\lambda_+ - \lambda_-}} \qquad , \qquad \sin{\alpha} =
\sqrt{\frac{a^2
- \lambda_-}{\lambda_+ - \lambda_-}} \:,\nonumber \\
                              \label{Seesawmixing}  \\
\cos{\beta} =\sqrt{\frac{\lambda_+ - b^2}{\lambda_+ - \lambda_-}}
\qquad , \qquad \sin{\beta} = \sqrt{\frac{b^2 -
\lambda_-}{\lambda_+ - \lambda_-}} \:.\nonumber \end{eqnarray}

\vspace{1mm}

\begin{eqnarray}  \cos{\alpha}\: \cos{\beta}=
\frac{c\:\sqrt{\lambda_+}}{\lambda_+ - \lambda_-} \quad , \quad
\cos{\alpha} \:\sin{\beta}=
\frac{b\:c^2\:\sqrt{\lambda_+}}{(\lambda_+ - b^2)(\lambda_+ -
\lambda_-)}        \nonumber \\
                             \\
\sin{\alpha} \:\sin{\beta}= \frac{c\:\sqrt{\lambda_-}}{\lambda_+ -
\lambda_-} \quad , \quad \sin{\alpha} \:\cos{\beta}=
\frac{a\:c^2\:\sqrt{\lambda_+}}{(\lambda_+ - a^2)(\lambda_+ -
\lambda_-)} \nonumber \end{eqnarray}

\vspace{3mm} \noindent Notice that in the space parameter $ a^2 \ll
c^2 \:,\:b^2 \; , \; \frac{\lambda_-}{\lambda_+} \ll 1$, so that we
may approach the eigenvalues as

\begin{equation} \lambda_- \approx \frac{D}{B} \approx \frac{a^2\:b^2}{c^2+b^2}
\qquad , \qquad \lambda_+ \approx c^2+b^2+a^2 -
\frac{a^2\:b^2}{c^2+b^2} \end{equation}

\end{document}